\documentclass[aps, pra, twocolumn, superscriptaddress, amsmath, showpacs, tightenlines, footinbib, longbibliography]{revtex4-1}
\usepackage[colorlinks, breaklinks, citecolor=blue, linkcolor=blue,  urlcolor={blue}]{hyperref}
\usepackage{breakurl}
\usepackage{amsfonts}
\usepackage{amsmath}
\usepackage{amssymb}
\usepackage{mathrsfs}
\usepackage{epsfig}
\usepackage{graphicx}
\usepackage[tight]{subfigure}
\usepackage{bm}

\usepackage{colortbl}
\definecolor{mygray}{gray}{.9}
\definecolor{darkblue}{rgb}{1,1,.70}
\definecolor{lightblue}{rgb}{1,1,.90}

\begin{document}

\title{Selective optomechanically-induced amplification with driven oscillators}
\author{Tian-Xiang Lu}
\affiliation{Key Laboratory of Low-Dimensional Quantum Structures and Quantum Control of Ministry of Education, Department of Physics and Synergetic Innovation Center for Quantum Effects and Applications, Hunan Normal University, Changsha 410081, China}
\author{Ya-Feng Jiao}
\affiliation{Key Laboratory of Low-Dimensional Quantum Structures and Quantum Control of Ministry of Education, Department of Physics and Synergetic Innovation Center for Quantum Effects and Applications, Hunan Normal University, Changsha 410081, China}
\author{Hui-Lai Zhang}
\affiliation{Key Laboratory of Low-Dimensional Quantum Structures and Quantum Control of Ministry of Education, Department of Physics and Synergetic Innovation Center for Quantum Effects and Applications, Hunan Normal University, Changsha 410081, China}
\author{Farhan Saif}
\affiliation{Department of Electronics, Quaid-i-Azam University, 45320 Islamabad, Pakistan}
\author{Hui Jing}\email{jinghui73@gmail.com}
\affiliation{Key Laboratory of Low-Dimensional Quantum Structures and Quantum Control of Ministry of Education, Department of Physics and Synergetic Innovation Center for Quantum Effects and Applications, Hunan Normal University, Changsha 410081, China}

\begin{abstract}
We study optomechanically-induced transparency (OMIT) in a compound system consisting of an optical cavity and an acoustic molecule, which features not only double OMIT peaks but also light advance. We find that by selectively driving one of the acoustic modes, OMIT peaks can be amplified either symmetrically or asymmetrically, accompanied by either significantly enhanced advance or a transition from advance to delay of the signal light. The sensitive impacts of the mechanical driving fields on the optical properties, including the signal transmission and its higher-order sidebands, are also revealed. Our results confirm that selective acoustic control of OMIT devices provides a versatile route to achieve multi-band optical modulations, weak-signal sensing, and coherent communications of light.
\end{abstract}
\maketitle

\section{Introduction} \label{Int}
Cavity optomechanics (COM), focusing on the interplay of optical lasers and mechanical devices,
provides unprecedented opportunities to explore both fundamental issues of quantum mechanics~\cite{Aspelmeyer86,Metcalfe1} and practical quantum control of light and sound~\cite{Bagci507,Teufel116,Korppi6,Gavartin7,Kampel7,Wollman349,Sillanpaa115,Lei117,Yamaguchi110,Grudinin104,Jing113,Jing8,ha1,ha2}. A prominent example, which is closely related to our present work, is optomechanically-induced transparency (OMIT)~\cite{wwu5,Kronwald111,Agarwal81,Weis330}. Playing a key role in COM-based coherent control of light, OMIT has been experimentally demonstrated with microtoroid resonators~\cite{Weis330}, diamond crystals~\cite{diamond}, microwave circuits~\cite{Teufel471}, nanobeam or membrane devices~\cite{Safavi472,Karuza88},  and nonlinear resonators~\cite{Dong1,Shen41}. In recent works, more exotic properties of OMIT devices have been revealed, such as cascaded OMIT~\cite{Fan6}, nonreciprocal OMIT~\cite{Shen10,Fang13,Shen,H5}, reversed OMIT~\cite{Jing5,Jing014006,Huilai}, vector OMIT~\cite{yingv}, nonlinear OMIT~\cite{Xiong86,Jiao18,Liu111}, two-color OMIT~\cite{Wang90}, and sub-Hertz OMIT~\cite{arxiv10184}. These devices provide a powerful platform to realize, for examples, quantum memory~\cite{Safavi472,Fiore107,Hill3}, signal sensing~\cite{Zhang86,wang91,Xiong42,Xiong95,Xiong110}, and phononic  engineering~\cite{Guo90,Ojanen90,Liu58,Zhang92}.

Very recently, COM devices fabricated with optical dimers (i.e., coupled optical resonators)~\cite{Tomita98,Grudinin104,zhangjing12,Xiaomin20,Peng101} or acoustic dimers~\cite{Fan6,Yamaguchi9,weaver,Lin4,Riedinger,Ockeloen}, have been utilized to achieve, for examples, COM-based phonon lasing~\cite{Grudinin104,Jing113,Jing8}, unconventional photon blockade~\cite{Li464,xiuminlin93,xiuminlin96,wu99}, and topological COM control~\cite{zhangjing12,harris,Yamaguchi1708}. In particular, by using multi-mode mechanical elements, experimentalists have demonstrated phonon-phonon entanglement~\cite{Riedinger,Ockeloen}, two-mode phonon laser~\cite{Yamaguchi110}, optomechanical Ising dynamics~\cite{Yamaguchi2}, mechanical synchronization or multi-wave phonon mixing~\cite{Lin4,Yamaguchi9,Colombano1810}, and coherent phonon transfer~\cite{arxiv09369}. Appealing predictions for this system also include acoustic Josephson junctions~\cite{BJJ}, COM superradiance~\cite{Agarwal,wuu99}, parity-time symmetry acoustics~\cite{XWXu92}, and phononic crystal shield~\cite{arxiv00561}.

\begin{figure}[ht]
	\centering
	\subfigure{
		\label{Fig1.sub.1}
		\includegraphics[width=7.3cm]{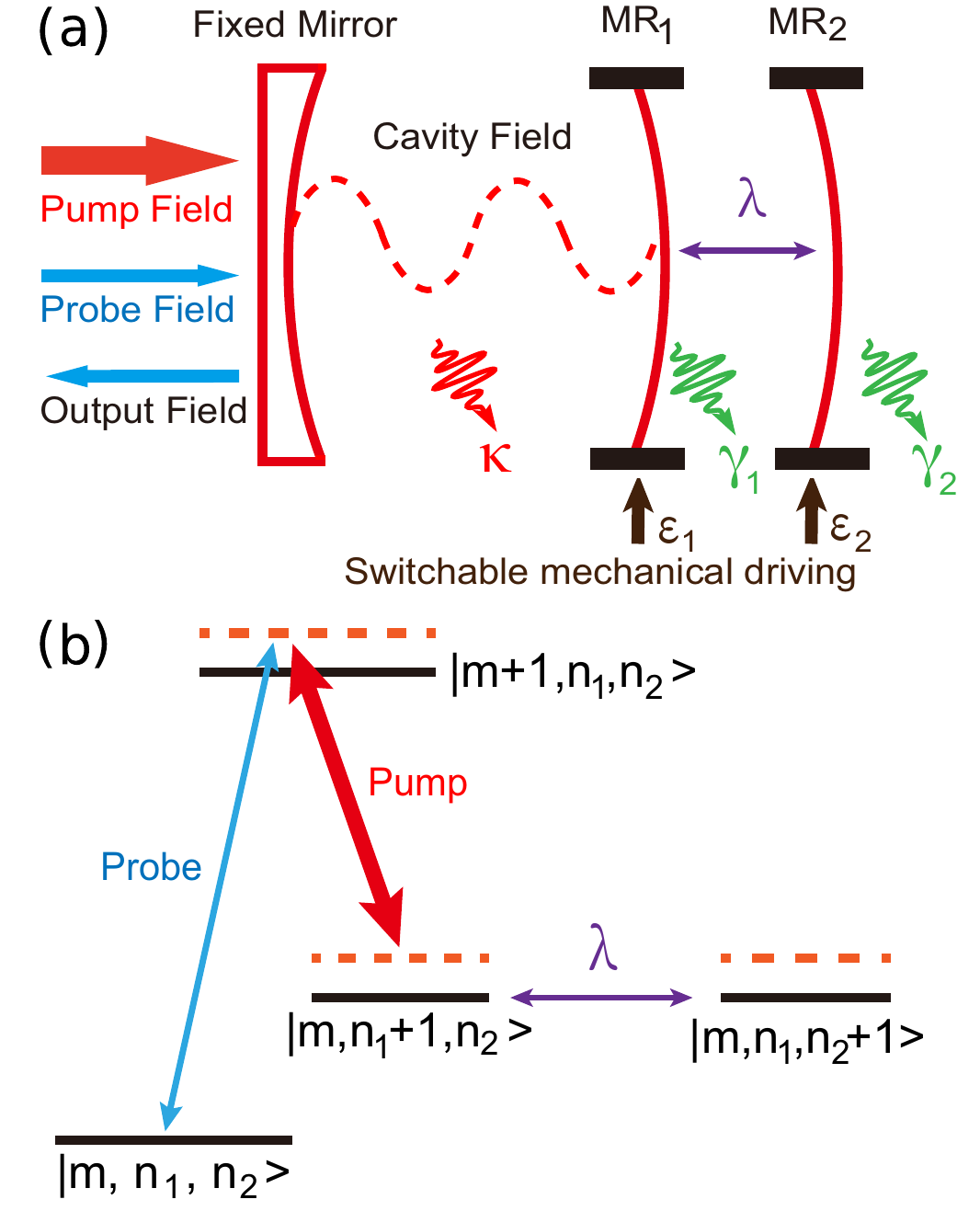}}	
		\caption{(Color online) (a) Schematic illustration of the compound COM system. The cavity is driven by a pump field at frequency $\omega_l$ and a weak probe field at frequency $\omega_p$, with the optical amplitudes $\varepsilon_l$, $\varepsilon_p$, and the phases $\phi_l$, $\phi_p$, respectively. The selective mechanical driving of MRs provides extra control of OMIT, with amplitude $\varepsilon_1$ ($\varepsilon_2$) at frequency $\omega_1$ ($\omega_2$). $\kappa$ and $\gamma_{i}$ $(i=1,2)$ are the optical and mechanical decay rates, respectively. (b) Energy-level structures of this system, where $|m\rangle$, $|n_1\rangle$ and $|n_2\rangle$ denote the number states of the cavity mode and the mechanical modes, respectively.}
	\label{Fig1}
\end{figure}
In this paper, we focus on the role of selective mechanical pump in OMIT with two coupled mechanical resonators (MRs). In experiments, this three-mode COM system has been demonstrated with a double-microdisk resonator, a zipper nanobeam photonic crystal, or a microwave device with two micromechanical beams~\cite{Massel3,Lin4}. Strong mechanical driving has also been utilized to achieve hybrid quantum spin-phonon devices~\cite{driving11} or ultra-strong exciton-phonon coupling~\cite{driving9}. In the absence of the mechanical driving, such a system features double OMIT spectrum, i.e., the appearance of two symmetric transparent peaks around an absorption dip at the cavity resonance (which is otherwise a transparent peak for COM with a single mechanical oscillator~\cite{Weis330,Safavi472}). Here we find that, by selectively driving the mechanical resonators, the OMIT peaks and the accompanied optical group delays can be significantly altered. In comparison with the case of only a single driven oscillator~\cite{Jia91,Xu92,liyong25,Si95,Suzuki92,JIANG,wu5}, in our system, we can achieve symmetric or asymmetric suppressions or amplifications of double OMIT peaks, which is accompanied by either significantly enhanced advance or a transition from advance to delay of the signal light. Our results confirm that multi-mode OMIT devices with selective acoustic control, provide a versatile route to realize coherent multi-band modulations, switchable signal amplifications, and COM based light communications.

\section{THEORETICAL MODEL}

We consider a three-mode COM system composed of an optical resonator with two MRs (see Fig.\,\ref{Fig1}). The MR$_1$ couples not only with the cavity field (via radiation pressure force), but also with the MR$_2$ (through the position-position coupling). As shown in experiments, the position-position coupling can be realized by e.g., using a piezoelectric transducer\cite{Yamaguchi9}, or applying an electrostatic force between the MRs~\cite{Tian,Hensinger}. For two MRs coupled by the Coulomb force~\cite{Tian,Hensinger,Ma90}, the interaction between them is written in the simple level as
\begin{align}
H_{\textrm{coul}}&=\frac{-k_{e}q_{1}q_{2}}{|r_{0}+x_1-x_2|}, \label{eq1}
\end{align}
where $k_{e}$ is the electrostatic constant, $r_{0}$ is the equilibrium separation of the two charged oscillators in absence of any interaction between them, and $m_i$ $(i=1,2)$ or $q_{i}=C_{i}V_{i}$ is the effective mass or the charge of the MR$_i$, with $C_{i}$ and $V_{i}$ being the capacitance and the voltage of the bias gate, respectively. $x_{i}$ $(i=1,2)$ is the small oscillations of the MR$_i$ from their equilibrium position. In the case of $r_{0}\gg \{x_{1},x_{2}\}$, with the second-order expansion, one can expand
\begin{align}
H_{\textrm{coul}}&\simeq\frac{-k_{e}q_{1}q_{2}}{r_{0}}\left[ 1- \frac{x_{1}-x_{2}}{r_{0}}+\left( \frac{x_{1}-x_{2}}{r_{0}}\right)^{2}\right], \label{eq2}
\end{align}
here the constant term and the linear term which can be absorbed into the definition of the equilibrium positions, and the quadratic term includes a renormalization of the oscillation frequency for the two MRs. Moreover, the small oscillations of the two mechanical oscillators from their equilibrium positions can be represented $x_{i}=\sqrt{\hbar/2m_{i}\omega _{m,i}}( b_{i}+b_{i}^{\dagger })$ $(i=1,2)$. Therefore, the effective Coulomb interaction can be simplified as
\begin{align}\label{}
\mathscr{H}_{\textrm{coul}}&=-\frac{2k_{e}C_{1}V_{1}C_{2}V_{2}}{r_{0}^{3}}x_{1}x_{2}=\hbar\lambda(b_{1}^{\dag}b_{2}+b_{1}b_{2}^{\dag}), \label{eq3}
\end{align}
here $\lambda$ is the Coulomb interaction strength
\begin{align}
\lambda=\frac{k_{e}C_{1}V_{1}C_{2}V_{2}}{r_{0}^{3}}\sqrt{\frac{\hbar}{{m_{1}m_{2}\omega _{m,1}\omega _{m,2}}}}\,,
\label{eq4}
\end{align}
for the typical experimental parameters $r_{0}=2\,\mathrm{mm}$, $C_{1}=C_{2}$ = 27.5$\,\mathrm{nF}$, and $V_{1}=V_{2} =1\,\mathrm{V}$~\cite{Tian,Hensinger}, we find $\lambda\simeq0.1\,\mathrm{MHz}$. Table~\ref{table} shows more relevant parameters of experimentally achieved coupled MRs. The cavity is driven by a pump field and a weak probe field. Meanwhile, as also shown in experiments~\cite{Fan6,Bochmann,bowenli}, mechanical driving fields with frequency $\omega_{i}$ and phase $\phi_i$ $(i=1,2)$ can be applied to selectively pump the MRs. We note that in the simplest two-mode COM (i.e., without the MR$_2$), pumping the MR$_1$ leads to a closed-loop $\Delta$-type energy-level structure [see Fig.\,\ref{Fig1}(b)], under which optical properties of the system become highly sensitive to the mechanical pump parameters~\cite{Jia91,Xu92,liyong25,Si95,Suzuki92,JIANG}. In the presence of coupled two MRs, as shown in a recent experiment~\cite{Fan6}, the effective phonon-phonon coupling also can be enhanced by the mechanical pump. Inspired by these works, here we show that by selectively driving the MRs, significantly different OMIT properties can be revealed, which offers flexible ways to control light in practice.
\begin{table*}[htbp]
\centering
\caption{\label{table}Experimental parameters of the mechanical resonators}
\begin{tabular}{p{17.5cm}}
 \rowcolor{darkblue}
 \ \ Reference {\qquad} {\qquad} {\qquad} \,Material {\qquad}\,\,\,\,\,\,Mechanical frequency $\omega_1(\omega_2)${\qquad}\,\,\,Damping rate $\gamma_1(\gamma_2)$ \,\,\,\,\,\,\,\,\,\,\,\,Coupling form\\
  \rowcolor{lightblue}
   \ \ L. Fan \textit{et al}.~\cite{Fan6}{\qquad}{\qquad}\,\,\,AlN {\qquad} {\qquad}\, $6.87\,\mathrm{GHz}\,(456.5\,\mathrm{MHz})${\qquad}{\qquad}\,\,\,\,\,\,\,\,\,\,\,$105.5\,\mathrm{kHz}\,(8.478\,\mathrm{kHz})$ \,\,\,\,\,\,\,\,\,\,\,nonlinearly \\
    \rowcolor{lightblue}
 \ \ Q. Lin \textit{et al}.~\cite{Lin4}{\qquad}{\qquad}\,\,\,$\mathrm{Si_3N}_4$ / $\mathrm{SiO}_2$\,\,\,\,\,\,\,$8.3\,\mathrm{MHz}\,(13.6\,\mathrm{MHz})$ {\qquad} {\qquad}{\qquad}\,$2.1\,\mathrm{MHz}\,(0.11\,\mathrm{MHz})$ \,\,~~~~~~\,linearly\\
  \rowcolor{lightblue}
 \ \ H. Okamoto \textit{et al}.~\cite{Yamaguchi9}\,\,\,\,\,\,~~~GaAs {\qquad}~~\,\,\,\,\,\,\,\,\,$1.845\,\mathrm{MHz}\,(1.848\,\mathrm{MHz})$ {\qquad}{\qquad}\,\,\,\,\,$131.9\,\mathrm{Hz}\,(131.9\,\mathrm{Hz})$\,\,~~~~~~\,\,~~linearly \\
  \rowcolor{lightblue}
  \ \ M. J. Weaver \textit{et al}.~\cite{weaver}\,\,\,\,\,\,~\,$\mathrm{Si_3N}_4${\qquad}~~\,\,\,\,\,\,\,\,\, $297\,\mathrm{kHz}\,(659\,\mathrm{kHz})$ {\qquad}{\qquad}{\qquad}~\,\,\,\,\,$9.42\,\mathrm{Hz}\,(6.28\,\mathrm{Hz})$\,\,~~~~~~~~\,\,\,\,\,\,\,linearly \\
\end{tabular}
\end{table*}

In the rotating frame at the pump frequency, the total Hamiltonian of the system can be written at the simplest level as
\begin{align}\label{}
H&=H_{\textrm{0}}+H_{\textrm{int}}+H_{\textrm{dr}},\nonumber \\
H_{\textrm{0}}&= \hbar\Delta _{c}c^{\dag }c+ \hbar\omega _{m,1}b_{1}^{\dag }b_{1}+
\hbar\omega _{m,2}b_{2}^{\dag }b_{2},  \nonumber \\
H_{\textrm{int}}&=- \hbar g c^{\dag }c\,(b_{1}^{\dag }+b_{1})+\hbar\lambda (b_{1}^{\dag }b_{2}+b_{1}b_{2}^{\dag }),\nonumber \\
H_{\textrm{dr}}&=i\hbar \sum\limits_{j=1,2}\varepsilon _{j}b_{j}^{\dag}e^{-i\omega _{j}t-i\phi_j}+i\hbar\varepsilon _{l}c^{\dag }\nonumber \\
&~~~+i\hbar\varepsilon_{p}c^{\dag }e^{-i\xi t-i\phi_{pl}}-\textrm{H}.\textrm{c}.,
\label{eq5}
\end{align}
where $c$ or $b_i$ $(i=1,2)$ is the annihilation operator of the cavity field with frequency $\omega_{c}$ or the MR$_i$ with frequency $\omega _{m,i}$, respectively. $\Delta_{c}=\omega_{c}-\omega_{l}$,  $\phi_{pl}=\phi_p-\phi_l$, and $g$ denotes the optomechanical coupling coefficient. In the following, we focus on the features of OMIT by selectively driving the MRs, including the signal transmission, group delay, and its higher-order sidebands.

For this purpose, the equations of motion (EOM) of this system are
\begin{align}
\dot{c}&=-\left( i\Delta _{c}+\frac{\kappa}{2}\right)
c+ig(b_{1}^{\dag }+b_{1})c+\varepsilon _{l}+\varepsilon _{p}e^{-i\xi t-i\phi_{pl}},\nonumber\\
\dot{b}_{1} &=-\left( i\omega _{m,1}+\frac{\gamma _{1}}{2}
\right) b_{1}+igc^{\dag }c-i\lambda b_{2}+\varepsilon
_{1}e^{-i\xi t-i\phi_1}, \nonumber\\
\dot{b}_{2} &=-\left( i\omega _{m,2}+\frac{\gamma _{2}}{2}
\right) b_{2}-i\lambda b_{1}+\varepsilon _{2}e^{-i\xi t-i\phi_2},
\label{eq6}
\end{align}
$\kappa$ and $\gamma_{i}$ $(i=1,2)$ are the optical and mechanical decay rates, respectively. In the case of $\{\varepsilon_p,\varepsilon_i\}\ll\varepsilon_l$, we express the dynamical variables as the sum of their steady-state values and small fluctuations, i.e., $c=c_{s}+\delta c$ and $b=b_{i,s}+\delta b_i$ $(i=1,2)$. The steady-state values of the dynamical variables are
\begin{align}
c_{s} &=\frac{\varepsilon_{l}}{ i\Delta+\frac{\kappa}{2} }, \nonumber\\
b_{1,s}& =\frac{ig|c_{s}|^{2}-i\lambda b_{2,s}}{ i\omega_{m,1}+\frac{\gamma_{1}}{2} }, \nonumber\\
b_{2,s}&=\frac{-i\lambda b_{1,s}}{ i\omega_{m,2}+\frac{\gamma_{2}}{2} },
\label{eq7}
\end{align}
here $\Delta=\Delta_{c}-g(b_{1,s}^{\ast}+b_{1,s})$. To calculate the amplitudes of the first-order and second-order sidebands, we assume that the fluctuation terms $\delta a$ and $\delta b_i$ $(i=1,2)$ have the following forms~\cite{Xiong86}
\begin{align}
\delta c &={A}_1^{-}e^{-i\xi t}+{A}_1^{+}e^{i\xi t}+{A}_2^{-}e^{-2i\xi t}+{A}_2^{+}e^{2i\xi t}, \nonumber\\
\delta b_{1} &={B}_1^{-}e^{-i\xi t}+{B}_1^{+}e^{i\xi t}+{B}_2^{-}e^{-2i\xi t}+{B}_2^{+}e^{2i\xi t}, \nonumber\\
\delta b_{2} &=D_1^{-}e^{-i\xi t}+D_1^{+}e^{i\xi t}+D_2^{-}e^{-2i\xi t}+D_2^{+}e^{2i\xi t}. \label{eq8}
\end{align}
\begin{figure*}[ht]
	\centering
	\subfigure{
		\label{Fig2.sub.1}
		\includegraphics[width=16cm]{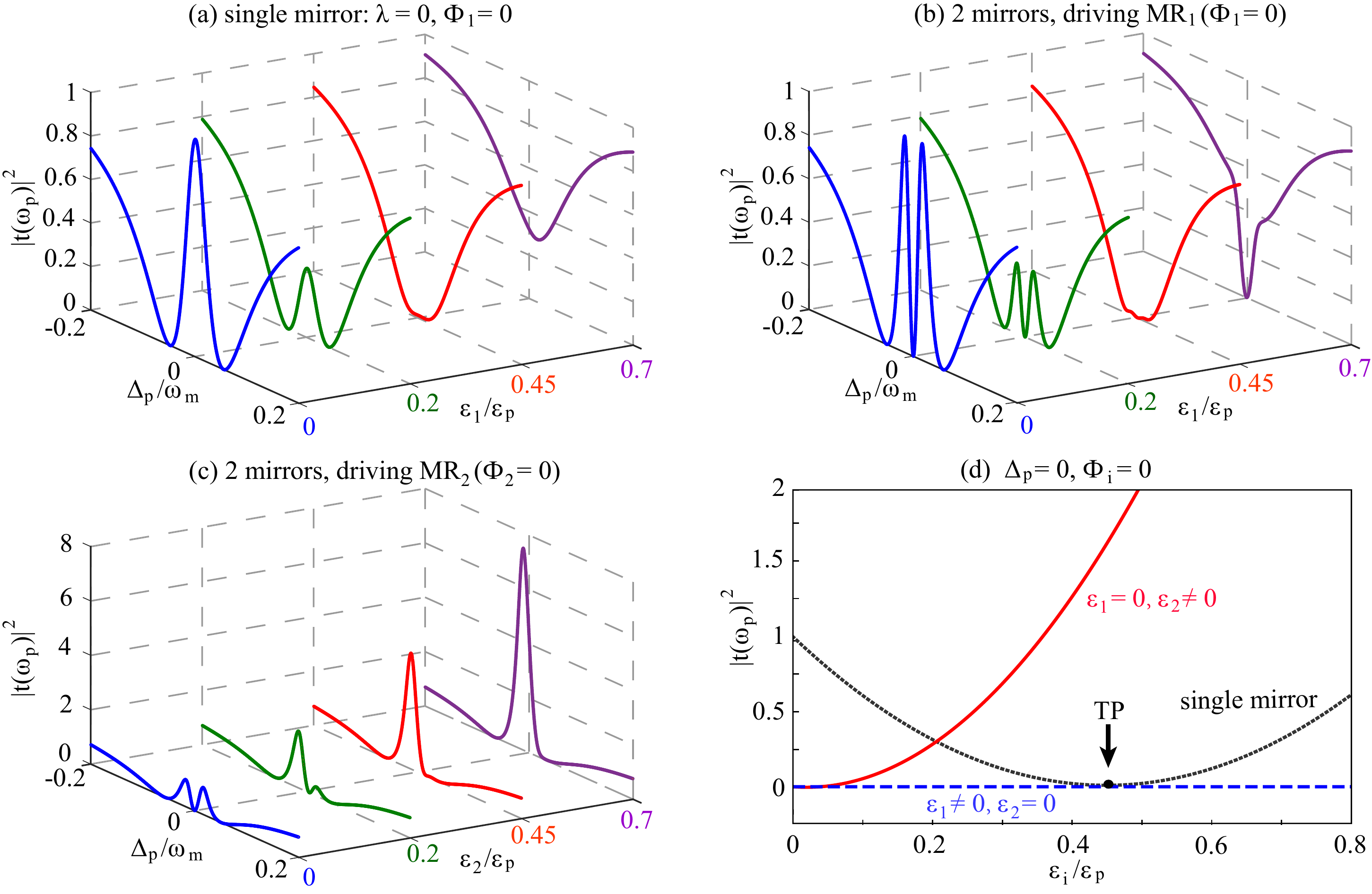}}
	\caption{(Color online) (a-c) Transmission of the probe light as a function of the optical detuning $\Delta_p/\omega_m$ with different values of the amplitude $\varepsilon_i$ $(i=1,2)$. (d) For $\Delta_p=0$, transmission of the probe light as a function of the amplitude $\varepsilon_i$ $(i=1,2)$.}
	\label{Fig2}
\end{figure*}

Substituting Eq.\,(\ref{eq8}) into Eq.\,(\ref{eq6}) leads to twelve equations. We can simplify these twelve equations into two groups~\cite{Xiong86}: one group describes the process of the first-order sideband which corresponds to the linear case
\begin{align}
h_{1}^+{A}_1^{-} &=iG({B}_1^{+\ast }+{B}_1^{-})+\varepsilon
_{p}e^{-i\phi_{pl}}, \nonumber\\
h_{1}^-{A}_1^{+\ast } &=-iG^{\ast }({B}_1{^-}+{B}_1^{+\ast }), \nonumber\\
h_{2}^+{B}_1^{-} &=iG{A}_1^{+\ast }+iG^{\ast }{A}_1^{-}-i\lambda D_1^{-}+\varepsilon _{1}e^{-i\phi_{1}}, \nonumber\\
h_{2}^-{B}_1^{+\ast }&=-iG^{\ast }{A}_1^{-}-iG{A}_1^{+\ast }+i\lambda D_1^{+\ast
}, \nonumber\\
h_{3}^+D_1^{-} &=-i\lambda {B}_1^{-}+\varepsilon _{2}e^{-i\phi_{2}}, \nonumber\\
h_{3}^-D_1^{+\ast } &=i\lambda {B}_1^{+\ast },   \label{eq9}
\end{align}
and the other group describes the the second-order sideband
\begin{align}
h_{4}^+{A}_{2}^{-} &=iG({B}_{2}^{+\ast}+{B}_{2}^{-})+ig(  {A}_{1}^{-}{B}_{1+}%
^{\ast}+{A}_{1}^{-}{B}_{1-}),  \nonumber\\
h_{4}^-{A}_{2}^{+\ast} &  =-iG^{\ast}({B}_{2}^{+\ast}+{B}_{2}^{-})-ig(
{A}_{1}^{+\ast}{B}_{1}^{-}+{A}_{1}^{+\ast}{B}_{1}^{+\ast}),  \nonumber\\
h_{5}^+{B}_{2}^{-}   &=iG{A}_{2}^{+\ast}+iG^{\ast}{A}_{2}^{-}+ig{A}_{1}^{-}{A}_{1}^{+\ast
}-i\lambda D_{2}^{-},\nonumber\\
h_{5}^-{B}_{2}^{+\ast}  &=-iG^{\ast}{A}_{2}^{-}-iG{A}_{2}^{+\ast}-ig{A}_{1}^{-}%
{A}_{1}^{+\ast}+i\lambda D_{2}^{+\ast},\nonumber\\
h_{6}^+D_{2}^{-}  & =-i\lambda {B}_{2}^{-},\nonumber\\
h_{6}^-D_{2}^{+\ast} & =i\lambda {B}_{2}^{+\ast},  \label{eq10}
\end{align}
here $G=gc_{s}$ and
\begin{align}\label{}
h_{1}^\pm=&\pm i\Delta+\frac{\kappa}{2}-i\xi,~~~h_{2}^\pm= \pm i\omega_{m,1}+\frac{\gamma_{1}}{2}-i\xi,\nonumber\\
h_{3}^\pm=&\pm i\omega_{m,2}+\frac{\gamma_{2}}{2}-i\xi,~~~h_{4}^\pm= \pm i\Delta+\frac{\kappa}{2}-2i\xi,\nonumber\\
h_{5}^\pm=& \pm i\omega_{m,1}+\frac{\gamma_{1}}{2}-2i\xi,~~~h_{6}^\pm= \pm i\omega_{m,2}+\frac{\gamma_{2}}{2}-2i\xi.\nonumber
\end{align}
By solving Eq.\,(\ref{eq9}) and Eq.\,(\ref{eq10}) leads to
\begin{widetext}
\begin{align}
{A}_1^{-} =\left[\frac{(h_{1}^-U_{1}^{+}U_{1}^{-}+|G|^{2}\Pi)\varepsilon _{p} }{h_{1}^+h_{1}^-U_{1}^{+}U_{1}^{-}+2i\Delta |G|^{2}\Pi}
+\frac{iGh_{1}^-h_{3}^+U_{1}^{-}\varepsilon _{1}e^{-i\Phi_{1}}}{h_{1}^+h_{1}^-U_{1}^{+}U_{1}^{-}+2i\Delta |G|^{2}\Pi}
+\frac{Gh_{1}^-U_{1}^{-}\lambda\varepsilon _{2}e^{-i\Phi_{2}}}{h_{1}^+h_{1}^-U_{1}^{+}U_{1}^{-}+2i\Delta |G|^{2}\Pi}\right]e^{-i\phi_{pl}},
\label{eq11}
\end{align}
and
\begin{align}
{A}_{2}^{-}=\frac{-i\xi gG\Gamma-gh_{4}^{-}U_{2}^{+}U_{2}^{-}(h_1^{-}/G^{\ast})}{h_{4}^{+}h_{4}^{-}U_{2}^{+}U_{2}^{-}+2i\Delta\left\vert G\right\vert ^{2}\Gamma  }A_{1}^{-}{A}_{1}^{+\ast}&+\frac{gG^{2}\Gamma (h_1^{-}/G^{\ast})}{h_{4}^{+}h_{4}^{-}U_{2}^{+}U_{2}^{-}+2i\Delta\left\vert G\right\vert ^{2}\Gamma}\left(  {A}_{1}^{+\ast}\right)^{2},
\label{eq12}
\end{align}
\end{widetext}
here
\begin{align}\label{}
\Phi_i=&~\phi_{i}-\phi_{pl}=\phi_i+\phi_l-\phi_p,\nonumber\\
U_{1}^\pm=&~h_{2}^\pm h_{3}^\pm+\lambda ^{2},~~~\Pi=h_{3}^-U_{1}^{+}-h_{3}^+U_{1}^{-},\nonumber\\
U_{2}^\pm=&~h_{5}^\pm h_{6}^\pm+\lambda ^{2},~~~\Gamma=U_{2}^{+}h_{6}^{-}-U_{2}^{-}h_{6}^{+}.\nonumber
\end{align}

With these at hand, by using the input-output relation~\cite{Collett} $$c_{}^{\textrm{out}}=c_{}^{\textrm{in}}-\sqrt{\eta_c\kappa}{A}_1^{-},$$ where $c^{\textrm{in}}$ and $c^{\textrm{out}}$ are the input and output field operators, respectively, we can obtain the transmission rate of the probe as
\begin{align}\label{}
|t(\omega_p)|^2=\left|1-\frac{\eta_c\kappa {A}_1^{-}}{\varepsilon_p e^{-i\phi_{pl}}}\right|^2. \label{eq13}
\end{align}
In Eq.\,(\ref{eq11}), the first term is the contribution from the standard OMIT process due to the destructive interference of the probe absorption~\cite{Weis330,Agarwal81}. The second and third term are the contribution from the phonon-photon parametric process~\cite{Jia91} and the phonon-phonon parametric process~\cite{Fan6}, induced by driving the MR$_1$ and MR$_2$. Clearly, these parametric process can modify and control the transmission of the signal field by adjusting the amplitude $\varepsilon_i$ and the photon-phonon mixing phase $\Phi_i$ $(i=1,2)$.

\begin{figure*}[ht]
	\centering
	\subfigure{
		\label{Fig3.sub.1}
		\includegraphics[width=17.5cm]{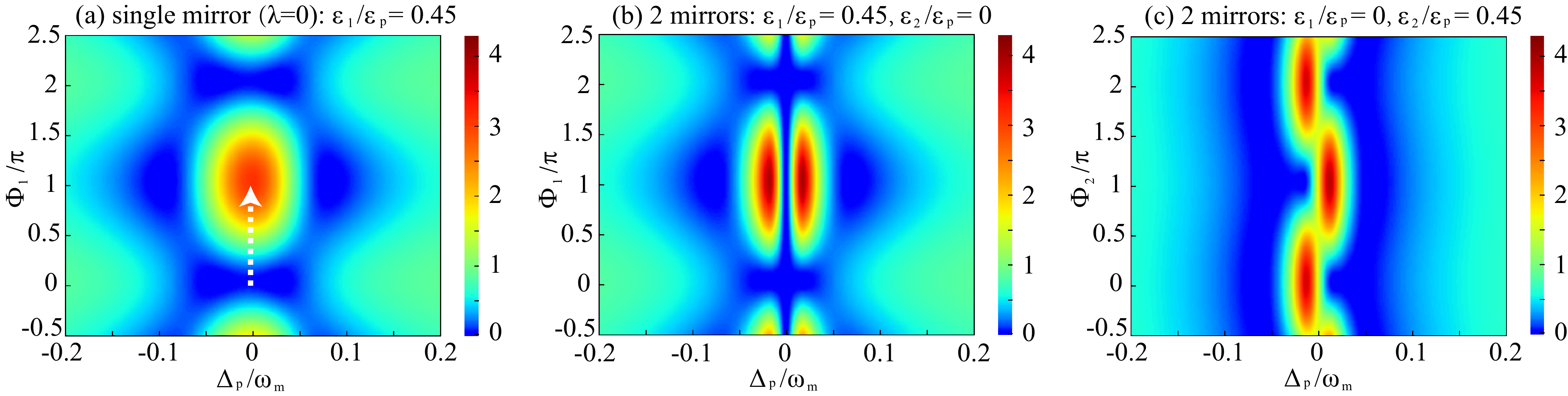}}	
	\caption{(Color online) (a) For single mirror ($\lambda=0$), the transmission as a function of the optical detuning $\Delta_p/\omega_m$ and the phase $\Phi_1$. (b,c) For two mirrors, transmission of the probe light as a function of the optical detuning $\Delta_p/\omega_m$ and the phase $\Phi_i$ $(i=1,2)$. }
	\label{Fig3}
\end{figure*}

\section{RESULTS AND DISCUSSION}
\subsection{Linear OMIT spectrum}
In our numerical simulations, to demonstrate that the observation of the signal transmission is within current experimental reach, we calculate Eq.\,(\ref{eq13}) and (\ref{eq17}) with parameters from Ref.~\cite{Hill3,weaver,Tian,Hensinger,Ma90}: $\omega _{m,i}/2\pi=947\,\mathrm{kHz}$ $(i=1,2)$, $m_{i}=145\,\mathrm{ng}$, $\gamma _{i}=\omega_{m,i}/Q$, $\kappa/2\pi=215\,\mathrm{kHz}$, $Q=6700$, $\lambda _{l}=1064\,\mathrm{nm}$, $L=25\,\mathrm{mm}$, $\lambda=0.1\,\mathrm{MHz},$ and $P_L=3\,\mathrm{mW}$. We have confirmed that for the pump power $P_L<7\,\mathrm{mW}$, single stable solution exists and the compound system has no bistability (see stability analysis in Appendix \ref{appendix A}).

Figure\,\ref{Fig2} shows the transmission rate $|t(\omega_p)|^2$ as a function of the optical detuning $\Delta_p/\omega_m=(\xi-\omega_m)/\omega_m$ and the phase $\Phi_1$. For comparisons, we first consider the single-mirror case ($\lambda=0$). As in standard COM system (without any mechanical driving), a standard single transparency window emerges around the resonance point $\Delta_p=0$ [see the blue solid line in Fig.\,\ref{Fig2}(a)], as a result of the destructive interference of two absorption channels of the probe photons (by the cavity or by the phonon mode)~\cite{Agarwal81,Weis330}.  When a mechanical driving field is applied to the MR$_1$, there are three coupling pathways of this system. The transitions $|m,n_1,n_2\rangle$ $\leftrightarrow$ $|m+1,n_1,n_2\rangle$, $|m,n_1+1,n_2\rangle$ $\leftrightarrow$ $|m+1,n_1,n_2\rangle$, and $|m,n_1,n_2\rangle$ $\leftrightarrow$ $|m,n_1+1,n_2\rangle$ can be achieved by applying a probe field, an optical pump field, and a mechanical pump field. Clearly, the three couplings result in a closed-loop $\Delta$-type transition strcture, leading to the phase-sensitive optical behaviors of the OMIT system~\cite{Jia91,Xu92,liyong25}. As shown in Fig.\,\ref{Fig2.sub.1},  the transmission rate at $\Delta_p=0$ can be firstly suppressed and then amplified by increasing the mechanical driving strength due to the interference between the OMIT process and the phonon-photon parametric process [represented by the first and the second terms in Eq.\,(\ref{eq11}), respectively]. By setting $|t(\omega_p)|^2=0$, the turning point (TP) position turns out to be
\begin{align}\label{}
\left(\frac{\varepsilon_{1}}{\varepsilon_{p}}\right)_{\textrm{TP}}&=\frac{\omega_{m,1}\kappa+\Delta\gamma_1
-\omega_{m,1}\eta\kappa\alpha}{2Gh_1^{+}h_1^{-}h_2^{+}h_2^{-}\eta\kappa}, \label{eq14}
\end{align}
with
\begin{align}\label{}
 \alpha=2h_1^{+}h_1^{-}h_2^{+}h_2^{-}+|G|^{2}\kappa\gamma_1-4\Delta\omega_{m,1}|G|^{2}, \label{eq15}
 \end{align}
 which, for the parameter values chosen here, corresponds to $({\varepsilon_{1}}/{\varepsilon_{p}})_{\textrm{TP}}\simeq0.45$.

For $\lambda\neq 0$, in the absence of any mechanical driving, double OMIT spectrum is known to appear in the two-mirror system [see the blue solid line in Fig.\,\ref{Fig2}(b)], i.e., the purely mechanical coupling splits the original single-mirror OMIT peak into two~\cite{wang91,Ma90}. Now we study the new features of double OMIT with switchable mechanical driving applied to either MR$_1$ or MR$_2$.

By driving the MR$_1$, both effective optoemchanical coupling and phonon-phonon coupling can be enhanced ~[27] and a closed-loop $\Delta$-type energy-level transitions configuration is formed in this system~(for similar systems, see Refs.~\cite{Jia91,Xu92,liyong25}). This leads to symmetric suppressions ($\Phi_1=0$) or  amplifications ($\Phi_1=\pi$) for both OMIT peaks [see Fig.\,\ref{Fig2}(b)], with a resonant absorption dip at $\Delta_p=0$ [Fig.\,\ref{Fig2}(b) and the blue dashed line in Fig.\,\ref{Fig2}(d)]. In contrast, by driving the MR$_2$, with the enhanced phonon-phonon coupling~\cite{Fan6}, highly asymmetric Fano-like OMIT spectrum appears due to the competition between the OMIT process and the phonon-phonon coupling process, corresponding to the first and the third terms in Eq.\,(\ref{eq11}), respectively, as shown in Fig.\,\ref{Fig2}(c). The physics of these features can be explained as follows: In such a system, the MR$_1$ couples not only with the cavity field, but also with the MR$_2$. By driving the MR$_1$, both effective optoemchanical coupling and phonon-phonon coupling can be enhanced~(see e.g., Ref.~\cite{Fan6} for similar results). However, by driving the MR$_2$, only effective phonon-phonon coupling can be enhanced. Thus, as expected, asymmetric amplifications of the signal light can be achieved by selectively driving the MR$_2$. These intuitive pictures agree well with our numerical results as shown in Fig.\,\ref{Fig2}.

 \begin{figure}[ht]
	\centering
	\subfigure{
		\label{Fig4.sub.1}
		\includegraphics[width=8cm]{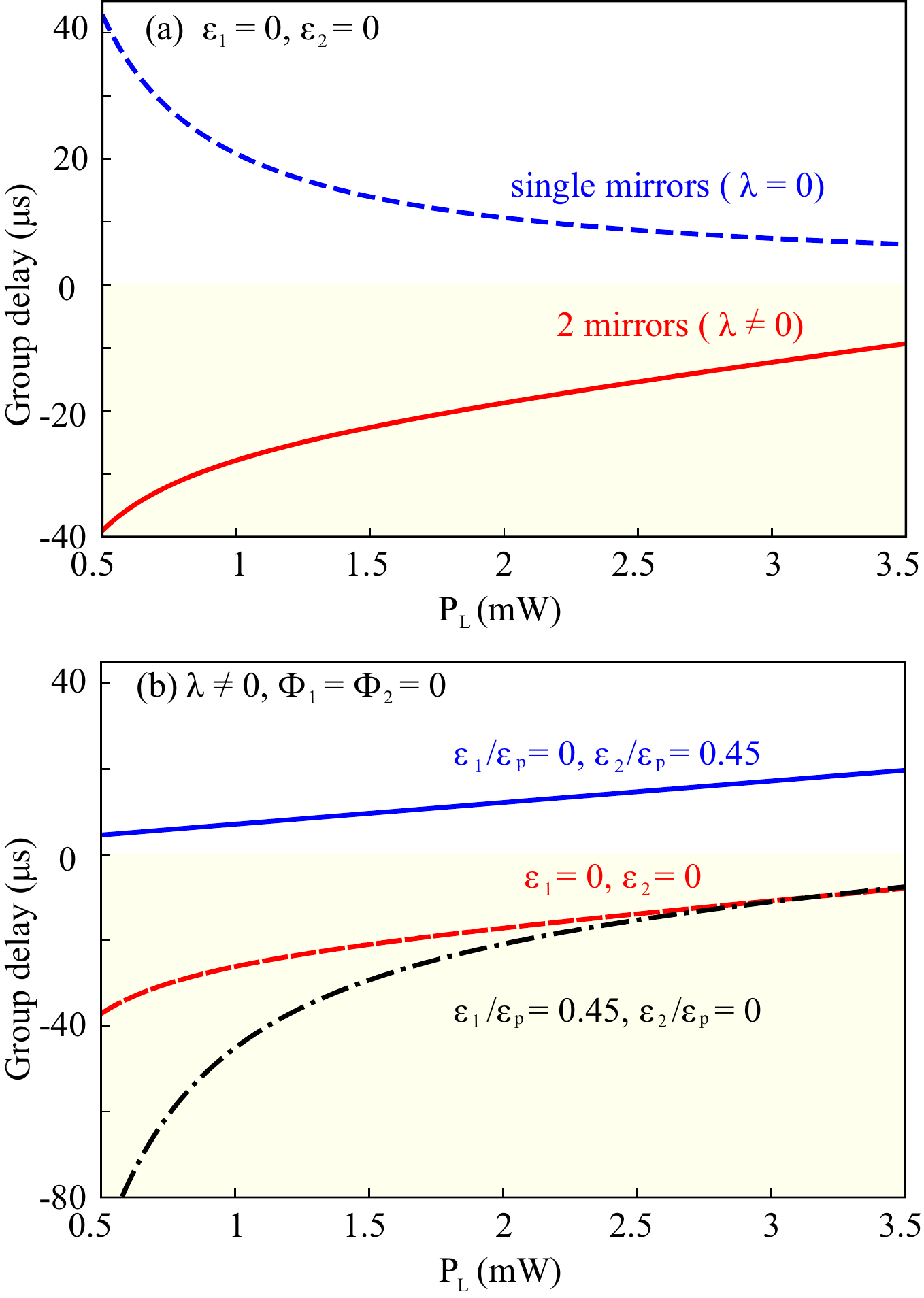}}
	\caption{(Color online) (a, b) For $\Delta_p=0$, group delay of the probe light $\tau_g$ (in the unit of $\mu s$) as a function of the pump power $P_L$.}
	\label{Fig4}
\end{figure}
\begin{figure*}[ht]
	\centering
	\subfigure{
		\label{Fig5.sub.1}
		\includegraphics[width=17cm]{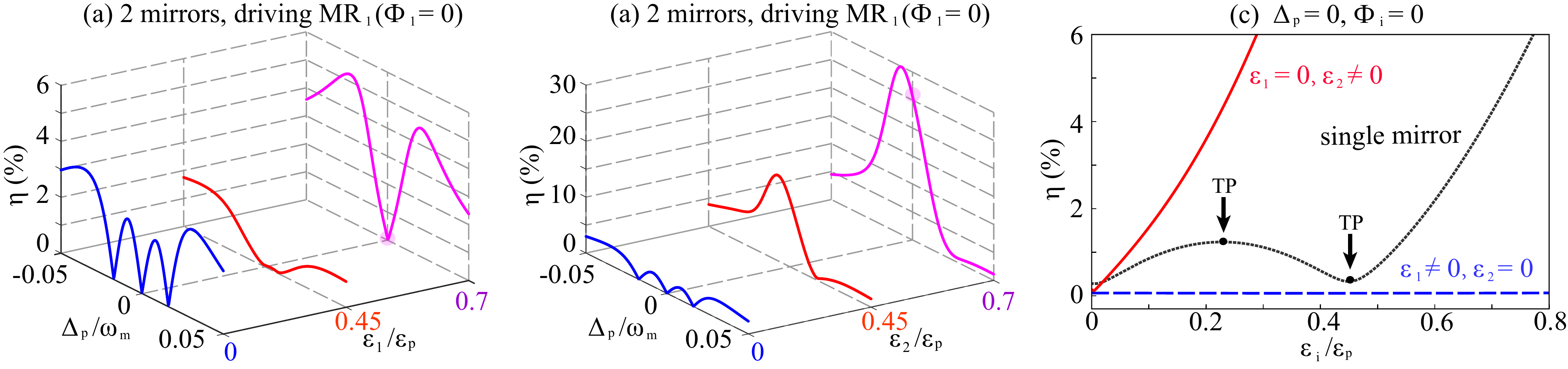}}
	\caption{(Color online) The efficiency of second-order sideband as a function of the optical detuning $\Delta_p/\omega_m$ with different values of the amplitude (a) $\varepsilon_1/\varepsilon_p$ and (b) $\varepsilon_2/\varepsilon_p$. (c) For $\Delta_p=0$, the efficiency of second-order sideband as a function of the amplitude $\varepsilon_i/\varepsilon_p$ $(i=1,2)$.}
	\label{Fig5}
\end{figure*}

Interestingly, for single mirror case, the transmission of the probe light changes periodically with the phase $\Phi_1$ [see Fig.\,\ref{Fig3}(a)]. For example, with $\varepsilon_1/\varepsilon_p=0.45$,  the transmission rate changes from strong absorption to amplification by tuning the phase $\Phi_1$ from 0 to $\pi$. Also, for two mirrors case, periodic changes of the optical transmission rate can be found by varying the phase of the mechanical driving field [see Fig.\,\ref{Fig3}(b) and Fig.\,\ref{Fig3}(c)]. Hence, more flexible OMIT control of the signal light becomes accessible by selective driving the MRs, e.g., the signal can be completely blockaded or greatly amplified by driving the MR$_1$ or MR$_2$, at the resonance point $\Delta_p=0$. This ability of selectively switching and amplifying the input weak signal could be highly desirable in practical optical communications~\cite{Jia91,Xu92,liyong25,Si95,Nunnenk}.

\subsection{Optical group delay}
The group delay of the transmitted light is given by
\begin{align}\label{}
\tau_g=\frac{d\arg[t(\omega_p)]}{d\omega_p}|_{\omega_p=\omega_c}. \label{eq16}
\end{align}
 Accompanying with the standard single-mirror OMIT, slow light [see the blue dashed line in Fig.\,\ref{Fig4.sub.1}] can emerge due to the abnormal dispersion~\cite{Safavi472}. In contrast, by introducing active gain into the system, fast light can be observed in experiments~\cite{Jing5,Xu92,wanglijun}. A merit of our system is the ability to selectively achieve either slow light or fast light by controlling the mechanical parameters. Figure\,\ref{Fig4}(b) shows that by driving the MR$_1$, significant enhancement of the light advance can be observed in comparison with the case without any mechanical pump [see the blue dashed line in Fig.\,\ref{Fig4}(b)]. However, by driving the MR$_2$, a tunable switch from fast to slow light can be achieved [see the blue solid line in Fig.\,\ref{Fig4}(b)]. For $P_L=3.5\,\mathrm{mW}$, in comparison with the single-mirror system, $\sim 5$ times enhancement can be observed for the group delay by using the two-mirror device. This is useful for achieving a multi-functional amplifier with the extra ability to selectively tune the optical group velocities.

\subsection{Nonlinear higher-order sidebands}
As defined in Ref.~\cite{Xiong86}, the efficiency of the second-order sideband process is
\begin{align}\label{}
\eta=\left| -\frac{\eta_c\kappa {A}_{2}^{-}}{\varepsilon_pe^{-i\phi_{pl}}}\right|. \label{eq17}
\end{align}
Due to nonlinear optoemchanical interactions, in the OMIT process, output fields with frequencies $\omega_l\pm n\xi$ can emerge, where $n$ is an integer representing the order of the sidebands~\cite{Xiong86}. The output fields with frequencies $\omega_l+2\xi$ is the second order upper sideband, while $\omega_l-2\xi$ is the lower sideband. In this work, we only consider the second-order upper sideband. For the single-mirror case, in the absence of any mechanical driving, the second-order sideband is subdued when the OMIT occurs, which results in a local minimum between the two sideband peaks around $\Delta_p=0$~\cite{Xiong86}. The efficiency of second-order sideband $\eta$ is, however, extremely small in conventional COM systems, e.g., $ 1\%-3\% $~\cite{Xiong86}.

As shown in Fig.\,\ref{Fig5}, by driving the MR$_1$, i.e., $\eta$ remains almost unchanged at the resonance [see Fig.\,\ref{Fig5}(a) and the blue dashed line in Fig.\,\ref{Fig5}(c)], which is similar to the linear OMIT spectrum [see the blue dashed line in Fig.\,\ref{Fig2}(d)]. In contrast, by driving the MR$_2$, giant enhancement of the second-order sideband can be observed at the resonance [see the red solid line in Fig.\,\ref{Fig5}(c)]. For example, for $\varepsilon_2/\varepsilon_p=0.7$, the efficiency $ \eta $ is about $25\%$ [see the purple solid line in Fig.\,\ref{Fig5}(b)], which is in sharp contrast to the corresponding result $\eta\approx0$ by driving MR$_1$. This giant enhancement of second-order sidebands, with much narrower bandwidth, can be used in precision measurement of very weak signals, e.g., single-charge detections~\cite{Xiong42,Xiong95}.

\section{Conclusion}
In conclusion, we have studied the mechanically controlled optical amplification and tunable group delay in a compound system composed of an optical resonator and two coupled mechanical resonators. We find that by driving one of the mechanical modes, both OMIT peaks can be symmetrically suppressed or amplified, which is accompanied by significantly enhanced light advance. In contrast, by driving the other mechanical mode, the OMIT spectrum becomes highly asymmetric, accompanied by a transition from fast light to slow light. In addition, periodic changes of both the linear OMIT spectrum and the higher-order sidebands can be observed by tuning the phases of the mechanical driving fields. These features of selective OMIT amplifications and switchable group delays of light provide more flexible ways in practical applications ranging from optical storage or modulations to multi-band optical communications. In future works, it will be also of interests to study the role of selective mechanical driving in enhancing or steering, for examples, light-sound entanglement~\cite{Vitali98,Nunnenkamp107}, photon-phonon mutual blockade~\cite{Rempe14}, precision measurement~\cite{Xiong42,Xiong95}, and switchable amplification of light or sound.

\textit{Note added}. After completing this work, we became aware of a preprint also on OMIT utilizing an acoustic dimer, but with only a fixed mechanical pump~\cite{arxiv1812}.

\section{ACKNOWLEDGMENTS}
This work is supported by the National Natural Science Foundation of China (NSFC) under Grants No. 11474087 and No. 11774086, and the HuNU Program for Talented Youth.

\appendix
\section{Stability analysis} \label{appendix A}
Considering photon damping and the Brownian noise from the cavity and the environment, the EOM are given by
\begin{align}
\dot{c}_{{}} &  =-\left(  i\Delta_{c}+\frac{\kappa}{2}\right)
c+ig(b_{1}^{\dag}+b_{1})c+\varepsilon_{l}+\sqrt{2\kappa}\,c_{\textrm{in}}\left(
t\right), \nonumber \\
\dot{b}_{1} &  =-\left(  i\omega_{m,1}+\frac{\gamma_{1}}{2}\right)
b_{1}+igc^{\dag}c-i\lambda b_{2}+\sqrt{2\gamma_{1}}\,\xi_{1}\left(  t\right), \nonumber \\
\dot{b}_{2} &  =-\left(  i\omega_{m,2}+\frac{\gamma_{2}}{2}\right)
b_{2}-i\lambda b_{1}+\sqrt{2\gamma_{2}}\,\xi_{2}\left(  t\right), \label{eqa1}
\end{align}
where $c_{\textrm{in}}\left(  t\right)  $ is the input noise operator with zero
mean value, and $\xi_{i}\left(  t\right)$ $(i=1,2)$ is the Brownian noise operators associated with the damping of the MR$_i$. Under the Markov approximation, two-time correlation functions of these input noise operators are
\begin{align}
\left\langle \hat{c}_{\textrm{in}}\left(  t\right)  \hat{c}_{\textrm{in}}\left(  t^{\prime}\right) \right\rangle  &  =\delta\left(  t-t^{\prime}\right),  \nonumber\\
\left\langle \xi_{i}\left(  t\right)  \xi_{i}\left(  t^{\prime}\right)  \right\rangle  &  =\left(  n_{\textrm{th}}+1\right)  \delta\left(  t-t^{\prime}\right)~~(i=1,~2),
\end{align}
here $n_{\textrm{th}}=\left(  e^{\hbar\omega/k_{B}T}-1\right) ^{-1}$, with $k_{B}$ is the Boltzmann constant and $T$ is the bath temperature. By setting all the time derivatives to zero of Eq.\,(\ref{eqa1}), the steady-state value of $c$ is
\begin{align}
c_{s} &  =\frac{\varepsilon_{l}}{\left(  i\Delta+\frac{\kappa}{2}\right) },
\label{eqa3}
\end{align}
where $\Delta=\Delta_{c}-g(b_{1,s}^{\dag}+b_{1,s})$ is the effective detuning,
including the effects of radiation pressure and Coulomb interaction. We now study the steady-state
behavior of the mean photon number $\left\vert c_{s}\right\vert ^{2}$ . In this case, using Eq.\,(\ref{eqa3}), it is straightforward to show that $\left\vert c_{s}\right\vert ^{2}$ satisfies
\begin{align}\label{}
\left\vert c_{s}\right\vert ^{2}\left(  \Delta^{2}+\frac{\kappa^{2}}{4}\right)  =\left\vert \varepsilon_{l}\right\vert ^{2}.
\label{eqa4}
\end{align}

We provide a direct and efficient estimation on how many positive solutions exist in Eq.\,(\ref{eqa4}) according to the Descartes rule. Eq.\,(\ref{eqa4}) can be recast as
\begin{align}\label{}
a_{3}x^{3}+a_{2}x^{2}+a_{1}x+a_{0}=0, \label{eqa5}
\end{align}
where we define $x=\left\vert c_{s}\right\vert ^{2}$, and the coefficients are
\begin{align}
a_{3}  & =W^{2}g^{4},~~~~~~~a_{2} =-2\Delta_{c}W g^{2},\nonumber\\
a_{1}  & =\frac{\kappa^{2}}{4}+\Delta_{c}^{2},~~~a_{0}=-\varepsilon_{l}^{2}, \label{eqa6}
\end{align}
with
\begin{align}
W=\frac{2\omega_{m,1}\left(  \omega_{m,2}^{2}+\frac{\gamma_{2}^{2}}
{4}\right)  -2\lambda^{2}\omega_{m,2}}{\left(  \omega_{m,1}^{2}+\frac
{\gamma_{1}^{2}}{4}\right)^2 -2\lambda^{2}\left(  \omega_{m,2}\omega_{m,1}-\frac{\gamma
_{1}\gamma_{2}}{4}\right)  +\lambda^{4}}, \nonumber\\
\end{align}
here all parameters $g$, $\kappa$, $\lambda$, $\omega_{m,1}$, $\omega
_{m,2}$, $\gamma_{1}$, $\gamma_{2}$, and $\varepsilon_{l}$ in Eq.\,(\ref{eqa6}) are
positive, we have $a_{0}<0$, $a_{1}>0$, $a_{2}<0$ and $a_{3}>0$, corresponding
to the following unique sign sequence:
\begin{align}
\text{sgn}\left(  a_{3}\right),...,\text{sgn}\left(  a_{0}\right)=+-+-.
\end{align}
\begin{figure}[ht]
	\centering
	\subfigure{
		\label{Figs2}
		\includegraphics[width=7cm]{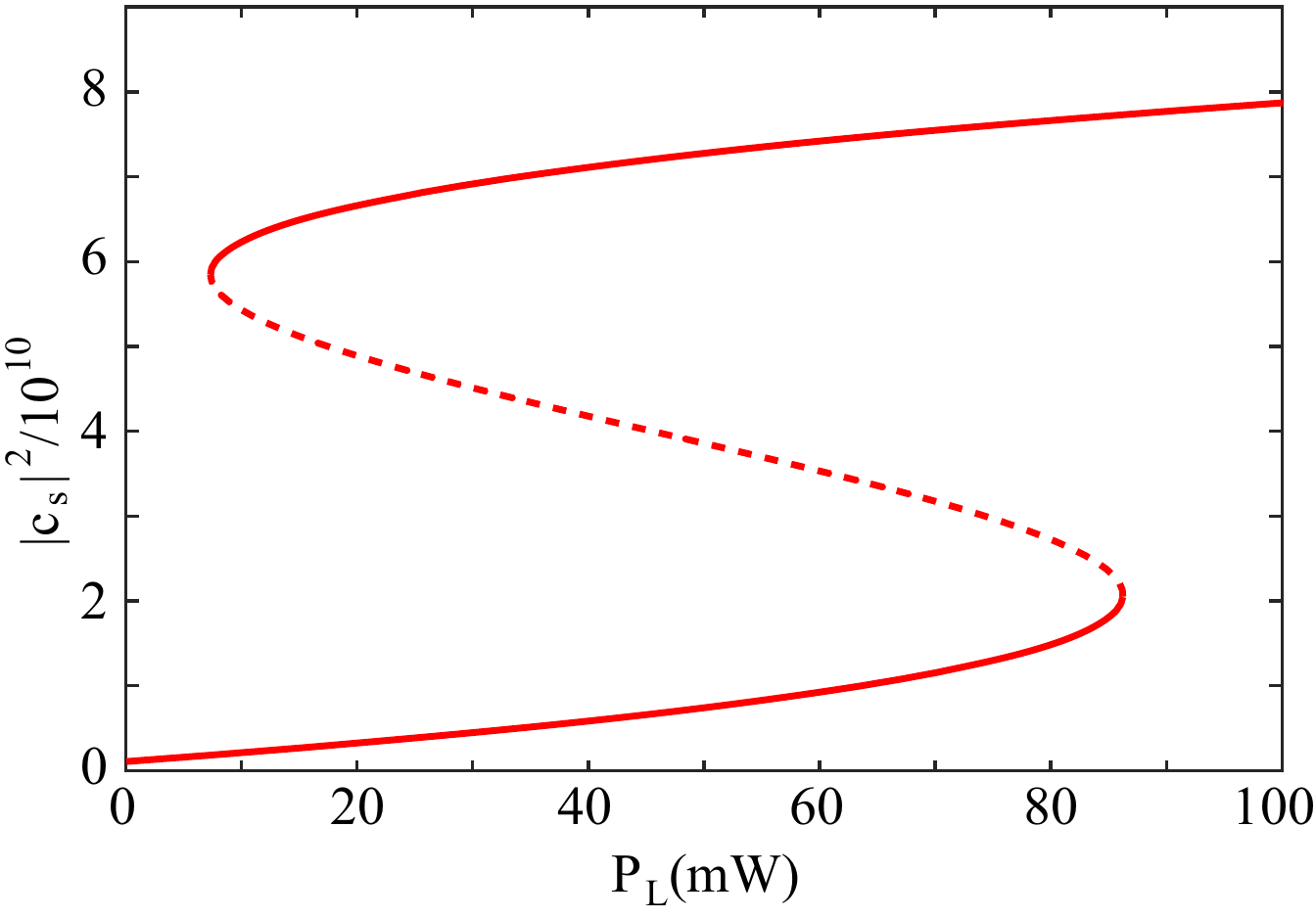}}
	\caption{(Color online) Mean intracavity photon number $\left\vert c_{s}\right\vert ^{2}$ as a function of the pump power $P_{L}$ with $\lambda=0.1\,\mathrm{MHz}$.}
\label{Figs2}
\end{figure}
According to the Descartes rule, Eq.\,(\ref{eqa5}) has three real solutions at most, two of which are dynamically stable. We also have checked numerically that the parameters we chosen in this paper satisfy the stability condition. When the cavity is driven on its red sideband, Figure\,\ref{Figs2} shows the mean intracavity photon number $\left\vert c_{s}\right\vert ^{2}$
as a function of pump power $P_{L}$ with $\lambda=0.1\,\mathrm{MHz}$. It can be seen that the
mean photon number exhibits the standard S-shaped bistability. As the pump
power $P_L$ increases from zero, there is only single stable solution of Eq.\,(\ref{eqa5}) at
the beginning. However, when $P_{L}$ is larger than a critical value, there
are three real solutions. The largest and smallest solutions are stable, and
the middle one is unstable.

Below we determine the stability of the steady states of our system using the
Routh-Hurwitz criterion~\cite{RH}. The fluctuation terms of the EOM are
\begin{align}
\delta\dot{c}&  =-\left(  i\Delta+\frac{\kappa}{2}\right)
\delta c+iG(\delta b_{1}^{\dag}+\delta b_{1})+\sqrt{2\kappa}\delta\hat{c}%
_{in}\left(  t\right),  \nonumber\\
\delta \dot{b_{1}} &  =-\left(  i\omega_{m,1}+\frac{\gamma_{1}}%
{2}\right)  \delta b_{1}+iG\delta c^{\dag}+iG^{\ast}\delta c-i\lambda\delta
b_{2} \nonumber \\
&~~~~+\sqrt{2\gamma_{1}}\delta\xi_{1}\left(  t\right),\nonumber \\
\delta\dot{b_{2}} &  =-\left(  i\omega_{m,2}+\frac{\gamma_{2}}%
{2}\right)  \delta b_{2}-i\lambda\delta b_{1}+\sqrt{2\gamma_{2}}\delta\xi
_{2}\left(  t\right),  \label{eqa9}
\end{align}
here $G=gc_{s}$, In a compact matrix form, Eq.\,(\ref{eqa9}) can be recast as
\begin{align}
\delta\dot{\textbf{u}}=\textbf{C}\textbf{u}+\delta \textbf{v}_{\text{in}}, \label{eqa10}
\end{align}
where vectors $\textbf{u}=(\delta c,\delta c^{\dag},\delta b_{1},\delta b_{1}^{\dag},\delta b_{2},\delta b_{2}^{\dag})^{\textrm{T}}$ and $\delta \textbf{v}_{\text{in}}=\sqrt{2}(\sqrt{\kappa}\,\delta\hat{c}_{in},\sqrt{\kappa}\delta\hat{c}_{in}{}^{\dag},\sqrt{\gamma_{1}}\,\delta\xi
_{1},\sqrt{\gamma_{1}}\delta\xi_{1}^{\dag},\sqrt{\gamma_{2}}\,\delta\xi
_{2},\sqrt{\gamma_{2}}\delta\xi_{2}^{\dag})  ^{\text{T}}$,  in which $\textrm{T}$
denotes the transpose of a matrix. The matrix $\textbf{C}$ is given by
\begin{align}
\textbf{C}=\left(
\begin{array}
[c]{cccccc}
-i\Delta-\frac{\kappa}{2} & 0 & iG & iG & 0 & 0\\
0 & i\Delta-\frac{\kappa}{2} & iG^{\ast} & iG^{\ast} & 0 & 0\\
-i\omega_{m,1}-\frac{\gamma_{1}}{2} & 0 & iG^{\ast} & iG & -i\lambda
& 0\\
0 & i\omega_{m,1}-\frac{\gamma_{1}}{2} & iG & iG^{\ast} & 0 &
i\lambda\\
-i\omega_{m,2}-\frac{\gamma_{2}}{2} & 0 & -i\lambda & 0 & 0 & 0\\
0 & i\omega_{m,2}-\frac{\gamma_{2}}{2} & 0 & i\lambda & 0 & 0 \\
\end{array}
\right). \nonumber
\end{align}
The characteristic equation $\left\vert \textbf{C}-\Upsilon \textbf{I}\right\vert =0$ can be
reduced to $\Upsilon^{6}+C_{1}\Upsilon^{5}+C_{2}\Upsilon^{4}+C_{3}\Upsilon
^{3}+C_{4}\Upsilon^{2}+C_{5}\Upsilon+C_{6}=0,$ where the coefficients can be
derived using straightforward but tedious algebra. From the Routh-Hurwitz
criterion~\cite{RH}, a solution is stable only if the real part of the
corresponding eigenvalue $\Upsilon$ is negative and the stability conditions
can then be obtained as

\begin{widetext}
\begin{align}
& C_{1}>0,\nonumber\\
& C_{1}C_{2}-C_{3}>0,\nonumber\\
& C_{1}C_{2}C_{3}+C_{1}C_{5}-C_{1}^{2}C_{4}-C_{3}^{2}>0,\nonumber\\
& C_{1}C_{2}C_{3}C_{4}+C_{2}C_{6}\left(  C_{1}^{2}+C_{3}\right)  +C_{1}
C_{5}\left(  C_{4}+C_{5}\right)  -C_{1}^{2}C_{4}^{2}-C_{1}C_{3}C_{6}-C_{3}
^{2}C_{4}-C_{4}^{2}>0,\nonumber\\
& C_{1}C_{2}C_{3}C_{4}C_{5}+\left(  C_{1}^{2}C_{2}-C_{2}C_{3}+C_{1}
C_{3}\right)  C_{5}C_{6}+\left(  C_{3}C_{2}+C_{1}C_{4}-C_{1}C_{2}^{2}
-C_{5}\right)  C_{5}^{2}-\left(  C_{1}C_{2}C_{6}+C_{4}C_{5}\right)C_{3}^{2}>0,\nonumber\\
& C_{1}C_{2}C_{3}C_{4}C_{5}C_{6}+\left(  C_{1}C_{4}^{2}-C_{1}^{2}C_{4}%
^{2}-C_{3}^{2}C_{4}\right)  C_{5}C_{6}+C_{2}C_{3}C_{5}^{2}C_{6}-C_{1}%
C_{2}C_{3}^{2}C_{6}^{2}-C_{1}C_{3}C_{5}C_{6}^{2}-C_{5}^{3}C_{6}>0. \label{eqa11}
\end{align}
\end{widetext}
Through these analyses, we have confirmed that the experimentally accessible parameters in the main manuscript can keep the compound system in a stable zone.

%\bibliography{nanomech.bib} %bibtex file
\end{document}